\documentclass[twocolumn,prb,showpacs]{revtex4}

\usepackage{graphicx}
\usepackage{rotating}
\usepackage{amsmath}
\usepackage{amsfonts}
\usepackage{amssymb}
\usepackage{enumerate}
\usepackage{longtable}
\usepackage{dcolumn}% Align table columns on decimal point
\usepackage{bm}

\begin{document}

\title{Hardcore Magnons in the $S=1/2$ Heisenberg Model on the Square Lattice}

\author{K.P. Schmidt$^1$}
\email{kaiphillip.schmidt@epfl.ch}
\homepage{http://marie.epfl.ch/~kpschmid}
\affiliation{$^1$Institute of Theoretical Physics, \'{E}cole Polytechnique
 F\'{e}d\'{e}rale de Lausanne, CH-1015 Lausanne, Switzerland}
\author{G.S. Uhrig$^2$}
\affiliation{$^2$Theoretische Physik, FR 7.1, Geb.\ E2.6, Universit\"at des 
Saarlandes, D-66123 Saarbr\"ucken, Germany}
\date{\rm\today}

\begin{abstract} 
We propose a versatile approach to treat commonly arising constraints.
It is illustrated for interacting magnons of the
Heisenberg antiferromagnet on a square lattice. 
For systems of $L\times L$ 
sites a non-perturbative continuous unitary transformation (CUT) is 
used to derive an effective Hamiltonian conserving the number of
magnons.  They are bosonic particles with a 
hardcore constraint which is captured  by a local, repulsive 
interaction $U$. The limits $U\to\infty$ and $L\to\infty$ are
achieved by  extrapolation. The residual spin gap  $\Delta_1$ is
smaller than $0.01J$ reflecting the gapless nature of the magnons. The 
one-magnon dispersion displays all known characteristics.
\end{abstract}


\maketitle

Strongly correlated systems in low dimensions display a large variety of 
fascinating properties like high-$T_{\rm c}$ superconductivity, 
quantum antiferromagnetism, quantum ferromagnetism or charge ordering. 
Theoretically, these  phenomena are investigated in simplified models like the
Hubbard model, the Heisenberg model and the $t$-$J$ model. These models are 
derived from ab-initio calculations, from more elaborate models, for instance
three-band Hubbard models, or they are motivated by 
phenomenological considerations. 

In the $t$-$J$ model transitions to
double occupancies are forbidden. This represents a hardcore constraint.
An analogous constraint arises if the magnons in an ordered Heisenberg model of
spin $S$ are treated as bosons. At most $2S$ magnons may be present at each 
site. A reliable treatment of such constraints is still a great challenge to
condensed matter theory. Straightforward approaches like mean-field 
approximations fail because of the  large (infinite)
energy scale of the hardcore repulsion.

In the present work we describe an innovative approach to treat hardcore 
constraints and other strong interactions. This approach is exemplified by the 
antiferromagnetic Heisenberg model on the square lattice. The ground state of 
this model displays long-range N\'eel order with a  sublattice 
magnetization $m\approx 0.3< 1/2$ reduced by quantum fluctuations. 
Therefore, a description in terms of magnons, propagating spin flips, as 
elementary excitations is a natural starting point which has inspired many
studies 
\cite{manou91,barne91,canal92a,hamer92,zheng93,klein95,singh95,sandv01,zheng05}.
These treatments are often based on expansions in the inverse
spin $1/S$ or in the inverse dimension $1/d$. So these approaches can be
problematic for low spin in two dimensions.

Magnons are interacting bosons subject to strong local constraints like
the hardcore property. The hardcore property has been dealt with by
choosing Hamiltonians with appropriate projections which ensure that
no unphysical states are reached in the course of the Hamiltonian dynamics
\cite{holst40,malee58,zitta65,schwi65}. Recent approaches replace the 
magnon  by other particles which do not exist freely or might exist only
at higher energies \cite{ho01,baska03}. The purpose of the present work
is to show that a direct treatment of the local constraint as an additional
interaction is possible and leads to very good results. This success opens
the route to treat also more complicated models like the $t$-$J$ model for
doped Mott insulators in an analogous way.

The Heisenberg model on the square lattice reads
\begin{equation}
\label{eq:hamilton-spin-definition}
 H = J \sum\limits_{\langle i,j\rangle} {\bf S}_{i}{\bf S}_{j}
\end{equation}
where $J>0$ denotes the antiferromagnetic exchange constant, ${\bf S}_{i}$ is 
the vector operator of a spin $S=1/2$ 
on site $i$, and $\langle i,j\rangle$ stands for the summation over 
nearest neighbors. We take the N\'eel state with $\downarrow$-spins on the 
$A$-lattice and $\uparrow$-spins on the $B$-lattice as  reference state. This
reference state will be mapped to the ground state by a continuous unitary
transformation (CUT). The elementary  excitations  are local spin 
flips which flip the $\downarrow$-spin up on the $A$-lattice or
the $\uparrow$-spin up on the $B$-lattice.
 Let $a^\dagger$ ($a$) be the corresponding usual bosonic creation 
(annihilation) operators. These magnon excitations are hardcore 
particles,  i.e., there can be at most one magnon per site. We impose this 
constraint by a local repulsive magnon-magnon interaction \cite{sushk98} 
\begin{equation}
 H^{U} = U \sum_{i} a^\dagger_i a^\dagger_i a^{\phantom{\dagger}}_i
 a^{\phantom{\dagger}}_i \ .
\end{equation} 
The hardcore constraint is recovered for $U\to \infty$.

In terms of the magnons, the Hamiltonian (\ref{eq:hamilton-spin-definition}) 
reads
\begin{eqnarray}
\label{eq:hamilton-magnon-definition}
 H^{\rm init}&=&\sum_{q}\left[ \omega_q a^\dagger_q a^{\phantom{\dagger}}_q+
B_q \left( a^\dagger_q a^\dagger_{-q}+ {\rm h.c.}\right)\right]
\nonumber \\
      &+& \sum_{k,k^\prime,q}
 V_{k,k^\prime,q}a^\dagger_{k^\prime+q}a^\dagger_{k-q}a^{\phantom{\dagger}}_k
 a^{\phantom{\dagger}}_{k^\prime}
\end{eqnarray} 
up to a constant. All energies are given in units of $J$.
The one-particle couplings are 
$\omega_q = 2$, $B_q = \frac{1}{2}\left( \cos (q_x) + \cos (q_y)\right)$,
 and the two-particle couplings are
$ V_{k,k^\prime,q} = U/N-1/(2N)\big[ \cos (k^\prime_x+q_x-k_x)+\cos
 (k^\prime_y+q_y-k_y)\phantom{\big]} + \phantom{\big[}\cos (q_x) + \cos
 (q_y)\big] = U/(2N)+f(k,k^\prime,q)$.
The Hamiltonian remains invariant under translations by one lattice constant 
$a$ (set to unity) in spite of the N\'eel state as reference state  because a 
spin rotation by 180$^\circ$ has been performed on one sublattice 
\cite{auerb94}. The two-particle couplings are manifestly symmetric under the 
exchange $k \leftrightarrow k'$ and $q \leftrightarrow k'+q-k$. 
Numerically,
we deal with a finite sample of linear size $L$ with $N=L\times L$ sites and
periodic boundary conditions.

A CUT \cite{wegne94,glaze93} 
%\cite{wegne94,glaze93,glaze94} 
is used to derive an effective Hamiltonian
$H^\text{eff}$ which conserves the number of magnons \cite{stein98,knett00a}. 
This is done by solving the flow equation
\begin{equation}
 \label{eq:flow}
 \partial_l H (\ell) = [\eta (\ell),H(\ell)]
\end{equation} 
where $\ell$ is a continuous auxiliary variable and 
$\eta(\ell)$ the infinitesimal  anti-Hermitian generator. 
The initial condition is $H(\ell=0)=H^{\rm init}$. 
In the present work, the CUT is realized in a self-similar, 
renormalizing fashion in momentum space. The commutators for all 
expressions are computed using the standard bosonic algebra in momentum space.
At maximum quartic terms in the bosonic operators are kept. Higher terms
involving six bosons are neglected after normal-ordering with respect to
the bosonic vacuum. So the Hamiltonian remains in the self-similar
form of a Hamiltonian of pairwise interacting bosons
\begin{eqnarray}
 H (\ell) &=&\sum_{q}\left[ \omega^{(\ell)}_q 
   a^\dagger_q a^{\phantom{\dagger}}_q+
 B^{(\ell)}_q \left(
 a^\dagger_q a^\dagger_{-q}+ {\rm h.c.}\right)\right]\\\nonumber
      &+& \sum_{k,k^\prime,q} V^{(\ell)}_{k,k^\prime,q}a^\dagger_{k^\prime+q}
 a^\dagger_{k-q}a^{\phantom{\dagger}}_k
 a^{\phantom{\dagger}}_{k^\prime}\\\nonumber
 &+& \sum_{k_1,k_2,k_3} \Delta^{(\ell)}_{k_1,k_2,k_3}\left(
 a^\dagger_{k_1}a^\dagger_{k_2}a^\dagger_{k_3}
 a^{\phantom{\dagger}}_{k_1+k_2+k_3}
 + {\rm h.c.}\right)\ .
\end{eqnarray}
The anti-Hermitian generator $\eta(\ell)$ is chosen to be
\begin{eqnarray}
 \eta (\ell) &=&\sum_{q} B^{(\ell)}_q 
 \left(a^\dagger_q a^\dagger_{-q}- {\rm h.c.}
\right)\\\nonumber
 &+& \sum_{k_1,k_2,k_3} \Delta^{(\ell)}_{k_1,k_2,k_3}\left(
 a^\dagger_{k_1}a^\dagger_{k_2}a^\dagger_{k_3}
a^{\phantom{\dagger}}_{k_1+k_2+k_3}
 - {\rm h.c.}\right)\quad .
\end{eqnarray}
The effective Hamiltonian $H^\text{eff}:=H(\ell=\infty)$ is characterized by 
$\omega_q^\text{eff}:=\omega_q^{(\ell=\infty)}$ and by
$ V^\text{eff}_{k,k^\prime,q} := V^{(\ell=\infty)}_{k,k^\prime,q}$; the
other terms have to vanish.

Comparing the coefficients of the same terms on the left and on the right
hand side of Eq.~\ref{eq:flow} yields the (high dimensional) 
set of differential equations to be solved, which is done numerically.
We treat systems of up to $14\times 14$ sites. 
The transformation can safely be 
carried out as long as the non-diagonal part of the Hamiltonian is decreasing 
monotonically \cite{reisc04} . As a measure of the  non-diagonal part we define
 the residual off-diagonality 
\begin{equation}
  \text{ROD}^2 := \frac{\sum_q
    {\big(B^{(l)}_q\big)^2}+\sum_{k_1,k_2,k_3}
    {\big(\Delta^{(l)}_{k_1,k_2,k_3}\big)^2}}{\sum_q
    {\big(B^{(l=0)}_q\big)^2}}\quad .
\end{equation}

Unfortunately, the numerical treatment of the flow equations reveals 
a significant increase of the ROD: no $H^\text{eff}$ can be obtained.
In looking for the reason for this failure we must keep in mind that
the ROD decreases as long as it is dominated by terms in which the change of 
the number of elementary excitations is correlated
to the change of the energy as measured by the diagonal part of
the Hamiltonian \cite{mielk98,knett00a,schmi05b}. In our system
this means that an increase in the number of magnons has to imply
an increase of the diagonal energy. If states with an incremented
number of excitations are lower in energy the CUT breaks down. So
a negative gap implies the failure of the CUT, see e.g.\
Ref.~\onlinecite{reisc04}. This happens also in the system at hand.
There is a bound state of two magnons with such a high binding energy that its 
excitation  energy is negative.

Why does this happen? The spin rotation symmetry requires that there
has to be a tightly bound state. Dealing with finite clusters there is
no true symmetry breaking so that the ground state is a singlet and
the elementary excitation is a gapped triplet with three degenerate states 
$S^z\in\{-1,0,1\}$. The description in terms of magnons, i.e.\
spin flips up or down, canonically provides
two of them, namely at $S^z\in\{-1,1\}$. The triplet with $S^z=0$ has to
be found in the sector with at least two magnons, one spin flip up and 
one down. The energy of this state is equal to the energy of an 
elementary magnon, i.e.\ the $S^z=0$ triplet must be a bound state of
two magnons. This argument unambiguously shows that a magnon description
of quantum antiferromagnets in or close to the paramagnetic phase 
 implies a very strong interaction between these magnons.

The fact that we neglected higher interactions  spoils
the delicate balance of magnon motion and magnon-magnon interaction. 
The numerics revealed that the attractive magnon-magnon interaction is 
overestimated leading to a too strongly bound $S^z=0$ state. To remedy 
this problem we reduce the interaction by hand (see below) 
so that the spin symmetry, namely the degeneracy between the one-magnon 
states at $\Delta_1:=\omega^\text{eff}_{q=(0,0)}$ and the two-magnon bound
state at $\Delta_2:=\omega^\text{2mag}_{K=(0,0)}$, is restored, 
i.e.\ $\Delta_1=\Delta_2$.
Here $\omega^\text{2mag}_K$ is the energy of the two-magnon bound 
state at total momentum $K$  which can be determined by standard numerics
once $H^\text{eff}$ has been found. For $K=(0,0)$,
we solve the secular equation
$H^\text{eff} \sum_q A_q |q\rangle|-q\rangle =\Delta_2
 \sum_q A_q |q\rangle|-q\rangle$ where 
$|q\rangle$ stands for a single magnon at momentum $q$.
The $S=1$  state with $S^z=0$ displays the symmetry $A_{q+(\pi,\pi)}=-A_q$.

The attractive interaction is deminished by reducing initial nearest-neighbor
attraction by the reduction factor $0<\lambda\le 1$
\begin{equation}
\label{eq:reduction}
 V_{k,k^\prime,q} \rightarrow V^{\lambda}_{k,k^\prime,q}=U/(2N)+\lambda
 f(k,k^\prime,q)\quad .
\end{equation}
The case $\lambda=0$ corresponds to the total omission of the attractive
magnon-magnon interaction keeping only the repulsive hardcore
interaction. The actual initial Hamiltonian $H^{\rm init}$ is recovered for
$\lambda=1$. The value of $\lambda$ is fixed to $\lambda_c$
where $\Delta_2=\Delta_1$ holds.
The above analysis of the failure of the direct CUT turns out to be valid.
As long as $\lambda \lessapprox \lambda_c$ the ROD decreases monotonically for large $\ell$ and the flow equations converge. 

The effective Hamiltonians $H^{\rm eff}(L,U)$ were obtained for
systems up to $L=14$ and $U=500$. The
flow equations were integrated until the ROD fell below  
$10^{-4}$. 
The large value of the interaction $U$ necessitates a careful numerical 
treatment of the flow equations for \emph{small} values of the flow 
$\ell\leq 1/U$  leading to a slowing down of the program. 
Thus parallelization is mandatory to be able to deal with large systems.

\begin{figure}[htbp]
  \begin{center}
    \includegraphics[width=\columnwidth]{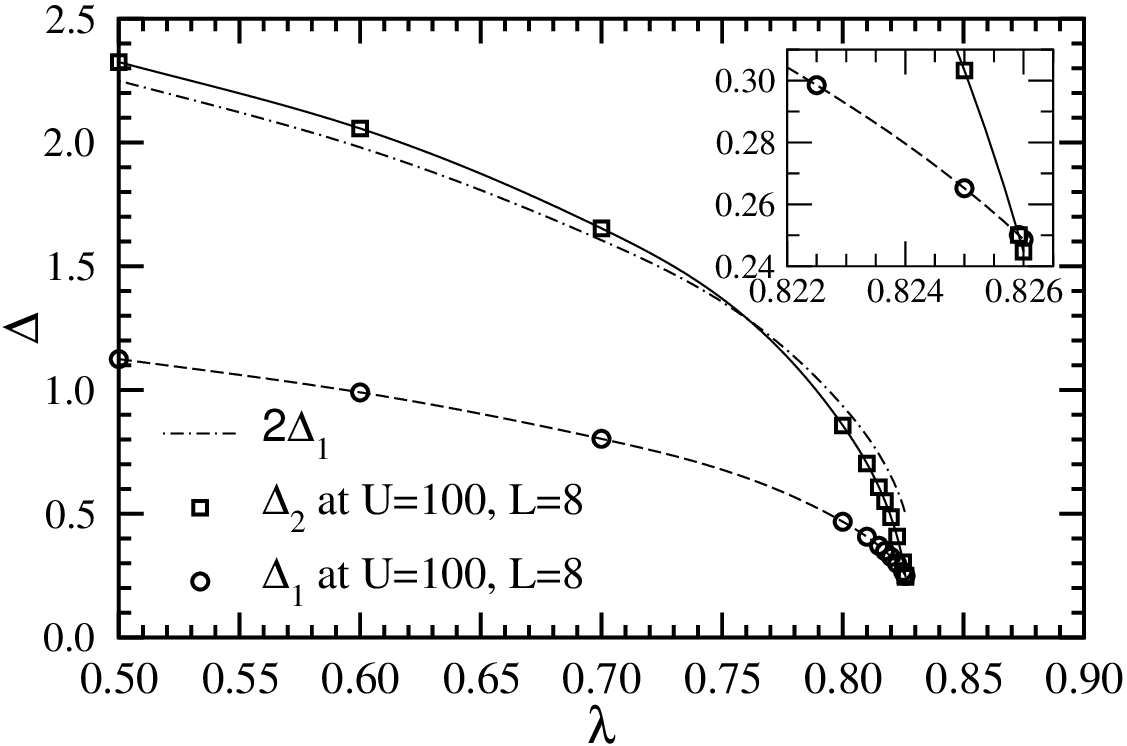}
  \end{center}
  \caption{Example of the one-magnon gap $\Delta_1$ (circles) and the 
    two-magnon bound 
    state energy $\Delta_2$ (squares) as function of 
    the  factor $\lambda$ in Eq.~\ref{eq:reduction}. 
    The solid lines are splines; the inset zooms at the intersection
    $\Delta_1=\Delta_2$.}
  \label{fig1}
\end{figure}

Fig.~\ref{fig1} depicts a generic example
 how the bound state energy $\Delta_2$ depends on 
the reduction factor $\lambda$. At $\lambda_c\approx 0.82592$ a  clear
intersection is discernible; the corresponding gap value is
$\Delta_1(L=8)=\Delta_2(L=8)=0.25008$. The fact that $\Delta_2 <
2\Delta_1$ (dashed-dotted curve in Fig.~\ref{fig1}) shows that
a true binding phenomenon is observed for appreciable
interaction $\lambda\gtrapprox 0.76$. For small values of $\lambda$ the 
repulsive interaction parametrized by $U$ dominates and 
the system shows a large gap of the order $J$.

\begin{figure}[htbp]
    \begin{center}
     \includegraphics[width=\columnwidth]{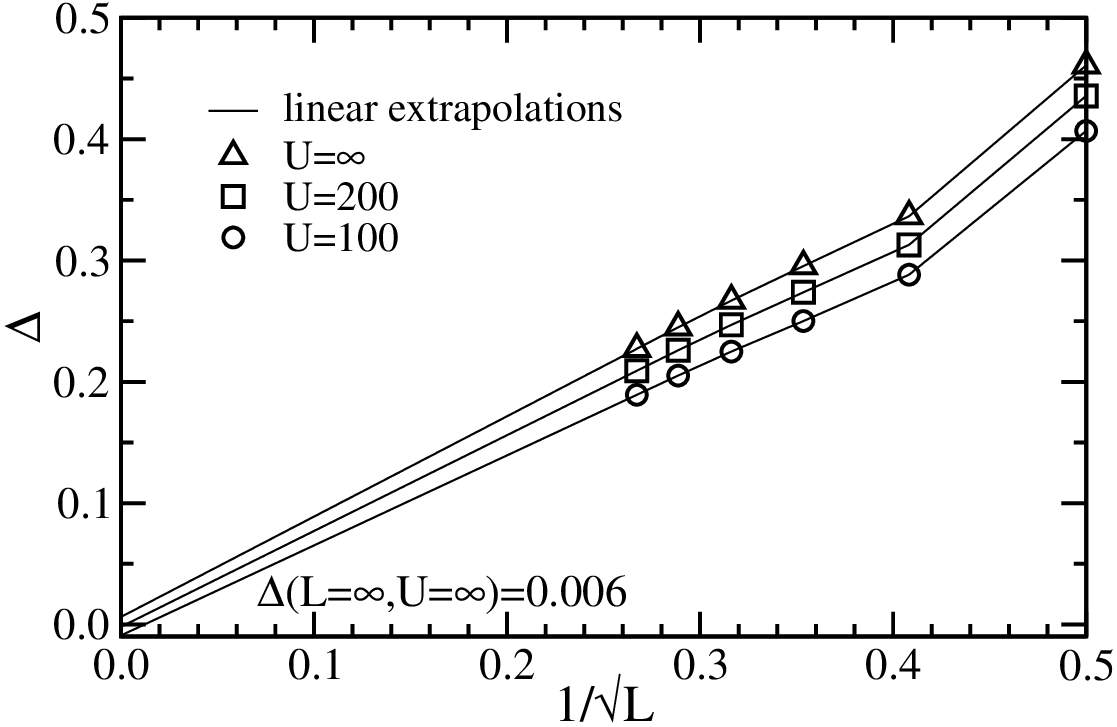}
    \end{center}
    \caption{The one-magnon gap $\Delta_1(L,U)$ as function of
      $1/\sqrt{L}$. The value at $U=\infty$ stem from linear 
      extrapolation in $1/U$. Solid lines are linear inter-/extrapolations.}
    \label{fig2}
\end{figure}

The same procedure  has been performed for many values of $L$ and $U$. 
All reduction factors $\lambda_{\rm c}$ range from $0.81$ to $0.83$.
Extrapolating them  in $1/U$ and in $1/L$ leads to 
$\lambda_c(L=\infty,U=\infty)= 0.825$. The deviation of 
$\lambda_c(L=\infty,U=\infty)$ from unity, i.e.\ $\approx 17\%$,
is a first estimate for the size of the truncation error.
In Fig.~\ref{fig2} the corresponding gaps $\Delta_1(L,U)$ are shown.
Linear extrapolation in $1/U$ yields the values at $U=\infty$. We found
that a subsequent extrapolation in $1/L$ does not work. But an
extrapolation in $1/\sqrt{L}$ works obviously very well. Except for $L=4$,
all data points lie nicely
on straight lines. This makes us confident to proceed
on the basis of the $1/\sqrt{L}$ extrapolation although we do  not know
of an {\it a priori} reason for this unusual scaling. We presume that
it results from using the N\'eel state as starting point for the treatment
of finite clusters.

It is reassuring that the residual gap $\Delta(L=\infty,U=\infty)$
is as small as $0.006J$. We take this fact as
evidence that we could restore the spin symmetry by the procedure of
reducing the initial attractive interaction in Eq.~(\ref{eq:reduction}).

\begin{figure}[htbp]
    \begin{center}
     \includegraphics[width=\columnwidth]{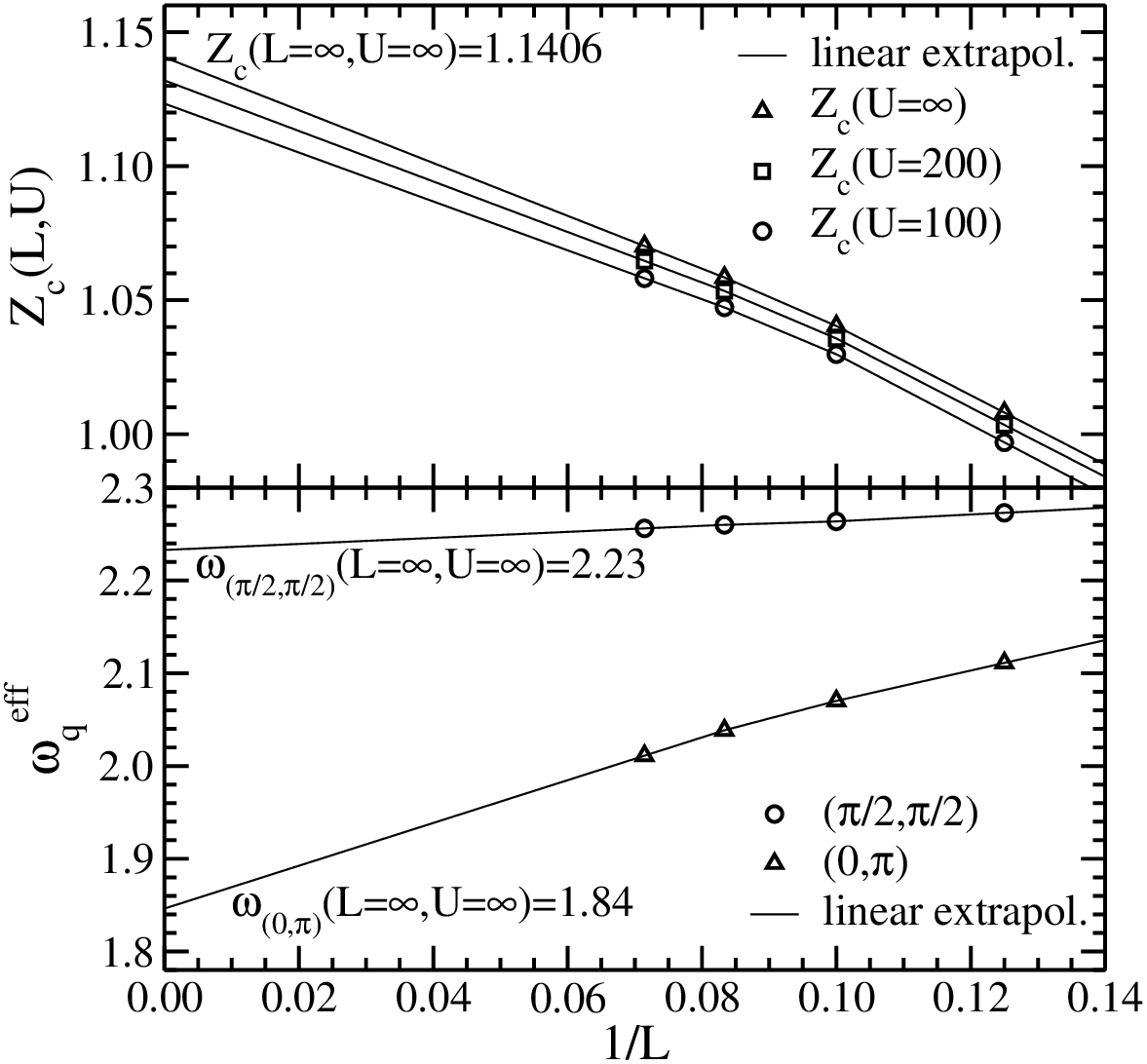}
    \end{center}
    \caption{Upper panel: quantum correction factor  $Z_c$ of the spin
      wave velocity is extrapolated in $1/U$ and $1/L$; for comparisons,
      see main text.
      Lower panel: inter-/extrapolation of the dispersion at two points
      of high symmetry.  The points represent results
      extrapolated linearly in $1/U$ to $U=\infty$.}
    \label{fig3}
\end{figure}

Next, we study the spin wave velocity $c$. It is characterized
by the correction factor $Z_c$ which quantifies the renormalization
relative to the result of spin wave theory $c=2\sqrt{2} Z_c$ 
\cite{manou91,canal92a,zheng93}. 
Since finite lattices are studied the determination of a group velocity at 
vanishing momentum is not possible. Thus we exploit the well-known shape of the
 dispersion 
\cite{manou91,auerb94} and derive the velocity from
\begin{equation}
c=\max_{q\in 1^\text{st} \text{BZ}} \omega_q^\text{eff}/|q|
\end{equation}
where we use implicitly that the spin gap $\Delta_1$ vanishes. 
The upper panel in Fig.~\ref{fig3} depicts the results and their 
extrapolations. Since the bosonic dispersion
$\omega_q^\text{eff}$ appears generically as square root 
$\sqrt{\Delta_1^2 + f(q)}$ with a smooth function $f(q)$ with $f((0,0))=0$
it is consistent to extrapolate finite values $\omega_q^\text{eff}$ 
in $1/L$ when the vanishing gap $\Delta_1$ scales like $1/\sqrt{L}$.
We obtain $Z_s(L=\infty,U=\infty)=1.14$ which is about
3\% away from the best values  $Z_c=1.17947$ (third order spin wave theory 
\cite{canal92a,zheng93}) and $Z_c=1.178(1)$ 
(high order series expansion \cite{zheng05}). In view of the error estimate
of 17\% on the basis of $1-\lambda_c$ the deviation in the low energy
part of the spectrum is small and the result shows that the CUT approach
is a valid way to tackle hardcore constraints.

At high energies it is known that the difference between
$\omega_{(\pi/2,\pi/2)}^\text{eff}$ and $\omega_{(0,\pi)}^\text{eff}$ 
is difficult to find. The third order spin wave results hardly
display any  dispersion between  $q=(\pi/2,\pi/2)$ and $q=(0,\pi)$
\cite{zheng93,zheng05}. But series expansion and quantum Monte
Carlo  calculations clearly show an appreciable difference between
the dispersion at $q=(\pi/2,\pi/2)$ and $q=(0,\pi)$
\cite{singh95,sandv01,zheng05}. One finds at $q=(\pi/2,\pi/2)$ a saddle point 
with  $\omega_{(\pi/2,\pi/2)}^\text{eff}\approx 2.17$ and at $q=(0,\pi)$ the 
maximum  with $\omega_{(0,\pi)}^\text{eff}\approx 2.39$.  In the lower panel of
Fig.~\ref{fig3}  the corresponding results in our approach are displayed. The
appropriate extrapolations finally yield
$\omega_{(\pi/2,\pi/2)}^{\rm eff}\approx 1.84$ and 
$\omega_{(0,\pi)}^{\rm eff}\approx 2.23$.
Hence, the former value deviates by about 15\% and the latter one by
about 7\%. These deviations do not surprise in view of the
first estimate of 17\% on the basis of $1-\lambda_c$. We conclude that the
omission of  terms beyond the two-magnon interaction
implies a truncation error of about 17\%. This is an encouraging
result because it is surely possible to include at least
the dominant parts of the higher magnon terms for further improvement.

Further support for the approach chosen comes from the 
fact that $\omega_{(\pi/2,\pi/2)}^{\rm eff}$ is significantly
lower than $\omega_{(0,\pi)}^{\rm eff}$ which agrees qualitatively
with the results by series expansion and quantum Monte Carlo. The
quantitative difference, however, is overestimated by about a factor
2 which is attributed to the truncation of higher magnon terms.
But the results show that the approach proposed captures the essential
physics also at higher physics in contrast to diagrammatic spin wave
theory.

In summary, we propose non-perturbative continuous unitary transformations
to treat the interactions arising from constraints.
This is successfully illustrated for magnons in the Heisenberg quantum
antiferromagnet on the square lattice. All qualitative aspects of the
dispersion are retrieved when the symmetry between elementary magnons and
bound pairs of magnons is restored. This finding demonstrates that
magnons in an antiferromagnet constitute a strongly correlated system.
Quantitatively, the results agree with previous findings  within 17\%.
Presently,  the CUT approach is less accurate than the highly
developed series expansions \cite{singh95,sandv01,zheng05}. 
But its accuracy can certainly be enhanced and it is a versatile tool which can
be used for the constraints in a multitude of models, including
also doped systems, because it is formulated in
the standard form of second quantization.

We thank A. L\"auchli, A. Reischl, S. Dusuel, F. Mila and E. 
M\"uller-Hartmann for stimulating discussions, the
Computer Centers of the Universit\"at zu K\"oln, the RWTH Aachen, and
 the FZ J\"ulich for computing ressources, and the DFG for funding in SFB 608.

%\bibliographystyle{apsrev}
%\bibliography{../bibinput/liter10}

\end{document}